\documentclass[prl,twocolumn,showpacs,superscriptaddress]{revtex4}
\usepackage{graphicx}

\begin{document}

\date{\today}

\title{Observation of the onset of strong scattering \\
on high frequency acoustic phonons in densified silica glass}

\author{Benoit Ruffl\'{e}}
\affiliation{Laboratoire des Verres, UMR 5587 CNRS, Universit\'{e} Montpellier 2, F-34095 Montpellier, France}
\author{Marie Foret}
\affiliation{Laboratoire des Verres, UMR 5587 CNRS, Universit\'{e} Montpellier 2, F-34095 Montpellier, France}
\author{Eric Courtens}
\affiliation{Laboratoire des Verres, UMR 5587 CNRS, Universit\'{e} Montpellier 2, F-34095 Montpellier, France}
\author{Ren\'{e} Vacher}
\affiliation{Laboratoire des Verres, UMR 5587 CNRS, Universit\'{e} Montpellier 2, F-34095 Montpellier, France}
\author{Giulio Monaco}
\affiliation{European Synchrotron Radiation Facility, Bo\^{\i}te Postale 220, F-38043 Grenoble, France}

\begin{abstract}
The linewidth of longitudinal acoustic waves in densified silica glass
is obtained by inelastic x-ray scattering.
It increases with a high power $\alpha$ of the
frequency up to a crossover where the waves experience strong
scattering.
We find that $\alpha$ is at least 4, and probably
larger.
Resonance and hybridization of acoustic waves with the boson-peak
modes seems to be a more likely explanation for these
findings than Rayleigh scattering from disorder.
\end{abstract}


\maketitle

A controversy regarding the fate of very high frequency
acoustic-like excitations in glasses arose in recent years \cite{Cou01,Ruo01}.
Acoustic waves propagate in glasses
up to rather high frequencies, for example up to at least
$\Omega/2\pi \simeq 0.4$ THz in vitreous silica, $v$-SiO$_2$ \cite{Rot84,Zhu91}.
However, silica shows a pronounced ``plateau'' in the temperature ($T$)
dependence of its thermal conductivity $\kappa(T)$ \cite{Zel71}.
This feature was explained by postulating the existence
of a rapid {\em crossover} of acoustic waves into a regime where they
experience {\em strong scattering} \cite{Gra86}.
In $v$-SiO$_2$, this crossover is expected at a frequency
$\Omega_{\rm co}/2 \pi \simeq 1$ THz \cite{Ray89}.
For the observed horizontal plateau to occur, besides strong scattering
of thermal phonons, phonons of lower $\Omega$ must also experience
a sufficiently strong $T$-independent scattering \cite{Ran88}.
Several models predict that the inhomogeneous width of acoustic waves can
increase with a high power of $\Omega$, typically in $\Omega^4$
\cite{Car61,Kar85}, leading to this rapid crossover.
However, the origin of the high power remained debated, whether Rayleigh
scattering from structural defects \cite{Gra86}, or resonance with local
modes \cite{Kar85,Dre68,Yu87}.
Rayleigh scattering seems generally too weak in materials as
homogeneous as dense glasses to be able to account for the relatively low
$\Omega_{\rm co}$ suggested by the position of the plateau \cite{Gra86,Ran88}.
The local modes are seen in the
``boson peak'' (BP) and produce an excess over the Debye
specific heat \cite{Buc86}.
Successful models based on resonance or on soft potentials
were developed to describe the observed
properties, {\em e.g.} in \cite{Kar85,Gra90,Par01}.

It might seem, judging from some recent literature \cite{Ruo01}, that the
above picture should be revised after it became
possible to observe in glasses the spectrum of longitudinal acoustic (LA) waves
at THz frequencies with x-ray Brillouin scattering (BS) \cite{Set98}.
This new experimental tool is important here,
as it potentially allows to check some of the previous conjectures.
It led to many publications claiming to prove that sound
{\em propagates} at frequencies much above the early expectations for
$\Omega_{\rm co}$, {\em e.g.} \cite{Ben96,Pil00}.
In this respect, it must be emphasized that inelastic x-ray scattering (IXS)
still is a difficult spectroscopy with severe limitations on
resolution and intensity.
Firstly, it is practically not possible with current instruments to
investigate {\em scattering} vectors $Q$ below $\sim 1$ nm$^{-1}$,
a value which happens to coincide with the expected {\em wave} vector $q$
at crossover, $q_{\rm co}$, in $v$-SiO$_2$ \cite{Ray89}
and in many other glasses.
Secondly, the narrowest instrumental profile allowing for sufficient intensity
\cite{Ver96} still has an energy
width around 1.5 meV (or $\simeq 0.4$ THz) and extended Lorentzian-like wings.
Owing to the relatively strong elastic scattering of glasses,
this tends to mask the weaker Brillouin signal.
For these reasons it is still not possible to investigate with IXS the LA
waves of $v$-SiO$_2$ at frequencies where their linewidth might grow
in $\Omega^4$ \cite{For96,Vac97}.
Finally, the signal-to-noise ratio being also small, the subtle changes in
BS lineshapes that indicate strong scattering might easily
go unnoticed \cite{For97}.
To alleviate several of these difficulties, we investigated
permanently densified silica glass, $d$-SiO$_2$, in which
$\Omega_{\rm co}/2\pi \simeq 2$ THz and  $q_{\rm co} \simeq 2$ nm$^{-1}$
\cite{Rat99,For02}.
A clear demonstration
of a rapid increase of the LA-width as $\Omega$ approaches
$\Omega_{\rm co}$ from below is still missing.
This is a {\em crucial} check for the onset of strong scattering.
It is all the more important that many
recent Letters advocated acoustic linewidths that vary
as $Q^2$ up to {\em very high} $Q$-values,
denying the onset of strong scattering and the existence of the
corresponding crossover, {\em e.g.}
\cite{Ben96,Pil00,Del98,Mon99,Ruo99,Ang00,Ruo00}.
In this Letter, we use the window between the smallest accessible
$Q$ and the $q_{\rm co}$ of $d$-SiO$_2$ to study the approach
of the crossover.
We find a width in $\Omega^{\alpha}$ with $\alpha$ at
least equal to 4 and probably larger, corroborating the
early intuition of a rapid crossover to
strong scattering.

\begin{figure}
\includegraphics[width=8.5cm]{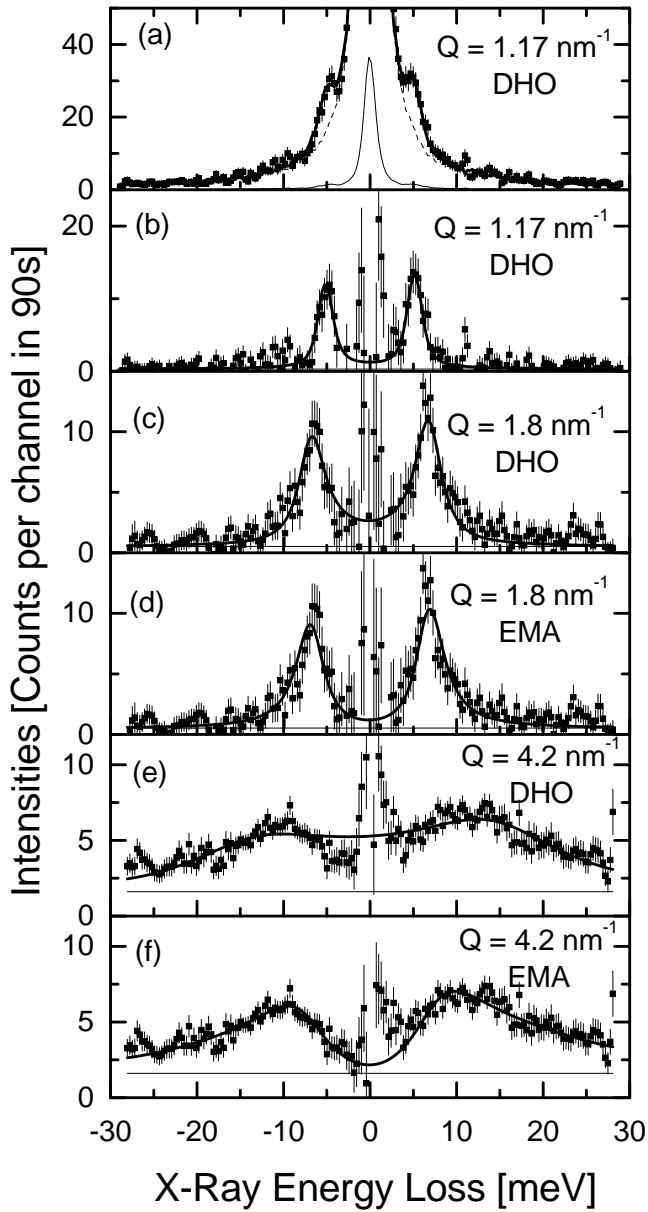}
\caption{IXS spectra of $d$-SiO$_2$ and their
fits explained in the text:
(a) A full spectrum and its central portion $\times 1/20$;
(b,c,e) DHO fits of the inelastic part after subtraction of
a CP freely adjusted in the fits;
(d,f) EMA fits of the inelastic part after subtraction of
a CP also freely adjusted.}
\end{figure}

The experiments were performed on the high-resolution IXS spectrometer
ID16 at the European Synchrotron Radiation Facility (ESRF) in Grenoble,
France.
The x-ray energy is 21.7 keV.
The main components of the spectrometer are a $T$-scanned monochromator,
the sample, a battery of five $T$-stabilized spherical analyzers, and the
corresponding detectors.
The instrumental lineshape was determined for each analyzer behind its
standard slit of
$20 \times 60$ mm (width$\times$height), using the signal from
the first sharp diffraction peak of polymethyl methacrylate at 20 K.
Very clean Voigt-like shapes were obtained with half-widths at
half-maximum of $\simeq 0.75$ meV.
The analyzers permit data collection at five $Q$
in increments of $\simeq 3$ nm$^{-1}$.
As we are now interested in small $Q$, the first analyzer
was placed to the left of the forward direction, and the second
to the right.
This allows simultaneous collection of two
interesting data points with $Q \le 2$ nm$^{-1}$.
The data presented here were thus acquired with only three different
positions of the first analyzer, $\mid Q \mid = 2, 1.8$, and 1.6
nm$^{-1}$.
Each spectrum was accumulated for nearly two full days.
The sample of $d$-SiO$_2$, of density 2.62 g/cm$^3$, is the same
as in previous measurements \cite{Rat99,For02}.
It is heated at 565 K to increase the inelastic signal.
At this $T$, the densified structure does not relax.
This was subsequently checked with optical BS,
the sound velocity providing
a sensitive test for the density \cite{Rat99}.

Figure 1 illustrates typical spectra and their fits.
The indicated values of $Q$ correspond to the center of the collection slit
which gives a spread $\Delta Q \simeq \pm 0.18$ nm$^{-1}$.
The spectra consist in a strong elastic peak plus a weaker Brillouin doublet.
In Fig. 1a, the full raw spectrum
is shown for the smallest usable $Q$, namely $Q=1.17$ nm$^{-1}$ \cite{foot}.
The dashed line follows the elastic peak, illustrating the
relative weakness of the doublet.
To extract information on the position $\Omega$ and half-width $\Gamma$,
the spectra are first adjusted to a damped harmonic oscillator (DHO) profile
plus an elastic central peak (CP), convoluted with the instrumental function.
Account is taken of the broadening due to the collection angle by
summing over contributions from surface elements of the slit,
each at its own $Q$, using $d\Omega /dQ = v_{\rm g}$.
Here $v_ {\rm g}$ is the group velocity which is determined iteratively.
There are four free parameters in the fits: the integrated intensities of
the elastic ($I_{\rm CP}$) and inelastic ($I_{\rm DHO}$) components,
$\Omega$ and $\Gamma$.
The small background shown by the baselines is fixed
to the detector noise, measured independently.
Figs. 1b,c illustrate, for the two extreme values of $Q$ measured below
crossover, the inelastic part that remains after subtraction of the adjusted
elastic contribution $I_{\rm CP}$.
The solid lines are the DHO fits.
In Fig. 1b the width is mostly instrumental, while in 1c it
is due to a large part to a real broadening of the Brillouin signal.

The values obtained for $\Gamma$ are shown in Fig. 2 against
those for $\Omega$ which agree with \cite{Rat99}.
$\Gamma$ increases very rapidly with $\Omega$.
There should be a homogeneous broadening contribution to this width,
$\Gamma_{\rm hom}$.
It is estimated from $\Gamma_{\rm hom} \propto \Omega^2$ by extrapolation
of an optical BS determination, $\Gamma_{\rm hom} / 2 \pi = 26 \pm 5$ MHz for
$\Omega / 2 \pi  = 41.5$ GHz at 600 K \cite{RatTh99}.
Interestingly, in that BS experiment the linewidths of $d$-SiO$_2$
are very near those of crystal quartz over a broad range of $T$,
whereas the linewidths of $v$-SiO$_2$ are typically four times larger.
Thus, it seems reasonable to assume $\Gamma_{\rm hom} \propto \Omega^2$, as
{\em e.g.} the Akhiezer mechanism might extend up to the THz region
in $d$-SiO$_2$ at these elevated $T$-value.
This gives the $\Gamma_{\rm hom}$ drawn in Fig. 2.
The total rate is then approximated by Matthiesen's rule of adding
rates due to independent processes,
$\Gamma = \Gamma_{\rm hom} + \Gamma_{\rm inh}$ \cite{Vac97}.
The inhomogeneous scattering width is adjusted to $\Gamma_{\rm inh} \propto
\Omega^\alpha$.
A fit of the five measured points with this expression gives the solid
line in Fig. 2 with $\alpha = 4.21 \pm 0.15$, and $\chi^2 = 0.76$.
For comparison, if we force $\alpha = 2$, the best fit shown by
the dotted line gives $\chi^2 = 5.4$.

\begin{figure}
\includegraphics[width=8.5cm]{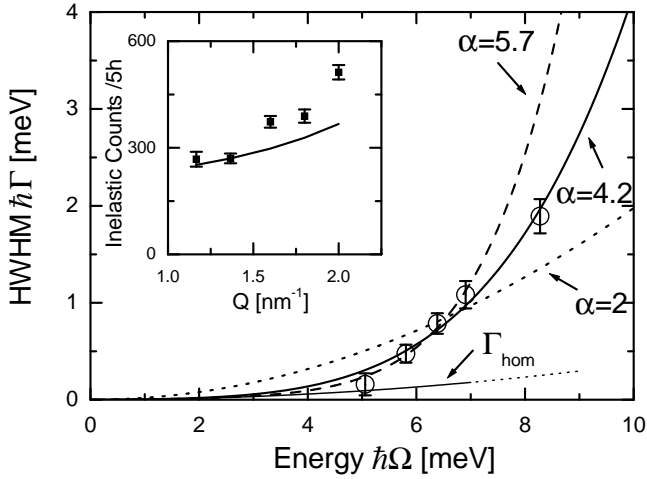}
\caption{The linewidths $\Gamma$ extracted from DHO fits.
$\Gamma_{\rm hom}$ extrapolated from optical Brillouin
scattering is shown by a full line terminated by dots.
A full line ($\alpha = 4.2$) shows the fit to $\Gamma_{\rm hom} +
\Gamma_{\rm inh}$.
The dashed line ($\alpha = 5.7$) is the same fit restricted to the four
lowest data points.
The dotted line illustrates the best fit with $\alpha = 2$, {\em i.e.} with
$\Omega^2$.
The inset shows the total inelastic intensities (dots) normalized to 5 hours
accumulation on the best analyser (\#2) compared to a line
$\propto v_{\rm LA}^{-2}$.}
\end{figure}

The inelastic intensities, $I_{\rm DHO}$, derived from these fits are
shown in the inset of Fig. 2.
With the very clean instrumental profile and the long accumulation times,
these are much more significant than previous ones \cite{Rat99}. 
They are compared to $v_{\rm LA}^{-2}$ (line), where
$v_{\rm LA} = \Omega / Q$ is
the phase velocity determined in this experiment.
The line is scaled to coincide with the most precise low $Q$ point.
One notes an excess in the observed $I_{\rm DHO}$ which grows as $Q$
approaches $q_{\rm co}$.
One should ask whether this is due to the use of the DHO instead of
the more appropriate effective-medium-approximation (EMA) lineshape
\cite{Rat99}.
To investigate this we adjust all spectra to the EMA, including
those obtained above $q_{\rm co}$ on the other detectors.
We find that $\hbar \Omega_{\rm co} = 8.5$ meV gives a good overall
adjustment with the other parameters as in \cite{Rat99}.
With these parameters fixed, there only remain the intensities
$I _{\rm EMA}$ and $I_{\rm CP}$ to fit each individual spectrum.
At small $Q$, for example for $Q=1.17$ nm$^{-1}$,
the EMA and the DHO are equivalent.
On the other hand, for $Q \gg q_{\rm co}$, the EMA is superior
as shown in Figs. 1e,f.
This particular spectrum is the average of three similar spectra acquired on
the third analyzer at $Q =$ 4, 4.2, and 4.4 nm$^{-1}$, for a total counting
time of nearly one week.
The three spectra were independently fitted and
then added, reducing the size of the error bars.
Fig. 1e shows that the DHO demands a larger signal at small $\omega$
than the EMA in Fig. 1f.
The fit adjusts this by using a smaller $I_{\rm CP}$ for the DHO, but this is
not able to represent properly the profile.
Below $q_{co}$, a similar comparison is shown in Figs. 1c,d.
One notices the same trend.
However, this is not the reason for the excess intensity.
Indeed, we find that $I_{\rm DHO} \simeq I_{\rm EMA}$ well within the
error bars for all five data points.
The small dip in the center of the EMA fits is compensated by their more
extended wings.
The intensity excess seems thus real.
For the present discussion, the main virtue of the EMA is to provide a fair
estimate for $\Omega_{\rm co}$.
There is of course no $\alpha$ to be extracted from this EMA
which {\em postulates} $\alpha=4$.

We now remark that the highest point in Fig. 2 is nearly at
$\Omega_{\rm co}$.
Owing to the Ioffe-Regel saturation of the damping, one cannot expect
that $\Gamma_{\rm inh} \propto \Omega^\alpha$ extends up to this point.
The saturation of $\Gamma_{\rm inh}$ is an important aspect of
the EMA \cite{Rat99}.
Therefore, it is more reasonable to only fit the lowest four points
to the power law.
The result is shown by the dashed line in Fig. 2, which corresponds
to $\alpha = 5.7 \pm 1.0$ with $\chi^2 = 0.28$.
The extrapolation of this fit to the highest point gives a $\Gamma_{\rm inh}$
nearly equal to twice the observed one.
This is precisely the saturation expected near crossover \cite{Rat99}.
The value of $\alpha$ obtained in this manner does not change
by allowing for the uncertainty in $\Gamma_{\rm hom}$, only its uncertainty
increases slightly.
If $\alpha$ is really so large, the onset of strong scattering of sound
cannot be due to Rayleigh scattering from structural inhomogeneities
which give $\alpha = 4$ \cite{Car61}.

\begin{figure}
\includegraphics[width=8.5cm]{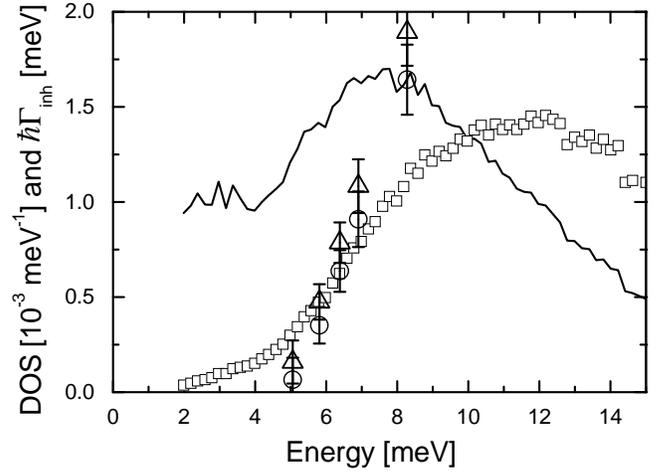}
\caption{The DOS of the local modes (squares) extracted
from neutron scattering data [33] compared to $\Gamma_{\rm inh}$
(dots) and $\Gamma$ (triangles). For illustration the boson peak calculated
from this DOS is shown by the full line in $10^{-5}$ meV$^{-3}$.}
\end{figure}

Resonance of sound with point-like local modes, assuming a bilinear
coupling, leads to $\Gamma _{\rm inh} (\Omega) \propto \rho(\Omega)$
where $\rho(\Omega)$ is the mode density of states (DOS) \cite{Kar85,Gra90}.
Fig. 3 shows the DOS of $d$-SiO$_2$ taken from \cite{Ina98} after
subtraction of a Debye contribution of $1.44 \times 10^{-5} \times
(\hbar \Omega)^2$, with $\hbar \Omega$ in meV.
The BP, $\rho(\Omega)/\Omega^2$, with its maximum near
$\Omega_{\rm co}$ \cite{For02}, is also shown.
The values of $\Gamma_{\rm inh}$ and $\Gamma$ are drawn on the same plot
for comparison.
They increase together with $\rho$ but do not seem exactly proportional to
$\rho$.
There can be several reasons for this.
One is that an exact determination of $\rho (\Omega)$ with neutron scattering
is always a difficult task, especially for $d$-SiO$_2$ as a large sample can
contain a few less densified fragments.
Another possible reason is more profound.
In silica the BP relates to rigid tetrahedra rotations \cite{Heh00}
which cannot be point-like in this connected network \cite{Tra02}.
Thus, it produces a diffuse optic branch which hybridizes with the acoustic
branch, an effect that depends both on the frequency and {\em size}
of the excitations.
Hence, there might be too few BP modes {\em of sufficient spatial
extent} (a few nm) to appreciably hybridize with acoustic phonons
at $q = 1.17$ nm$^{-1}$.
The anomalous BS intensity increase near $\Omega_{\rm co}$ can also
result from this hybridization.
Either the hybridization effectively softens the LA waves
increasing their fluctuation amplitudes over what $v_{\rm LA}$ suggests,
or the coupled local modes give
themselves a coherent contribution to the BS amplitude.
Further investigations are needed to clarify these points.

\begin{figure}
\includegraphics[width=8.5cm]{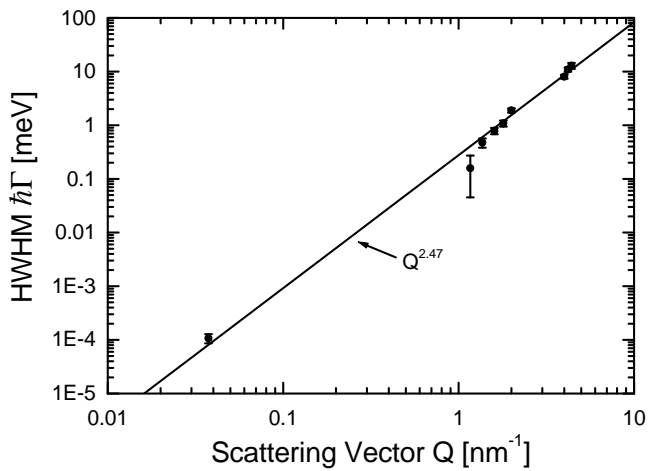}
\caption{Logarithmic presentation of $\Gamma (Q)$ for the optical
Brillouin scattering point [31] and the IXS data obtained here.
The straight line gives a power $\gamma = 2.47$ which averages over
different regimes and which is of no further significance.}
\end{figure}

As final remark we stress that a logarithmic presentation of
$\Gamma (Q)$ over a large range of $Q$ can be misleading.
This is illustrated in Fig. 4 for the data presented here.
The line is a fit with $\Gamma \propto Q^\gamma$, which gives
$\gamma = 2.47 \pm 0.08$.
Such plots have often been taken as evidence for the absence of
crossover phenomena, {\em e.g.} in \cite{Ben96,Ruo99,Ruo01}.
The above discussion shows that this reasoning is unfounded.
In Fig. 4, the lowest $Q$ point is purely homogeneously broadened, the
next four points are dominated by inhomogeneous broadening, the
sixth point is near crossover, and the following three points are well
above crossover.
Hence, that presentation averages over three distinct regimes:
one where $\Gamma = \Gamma_{\rm hom}$, one which is dominated by 
$\Gamma_{\rm inh}$, and finally one where $\Gamma$ should not be extracted
from DHO fits.
The rapidly growing $\Gamma_{\rm inh}$ only dominates $\Gamma$ over a
narrow range in $\Omega$ (or $Q$).
To be able to conclude about $\alpha$ as
we do here, one needs: $i$) to know the position of $\Omega_{\rm co}$
and, $ii$) to have a sufficient number of data points with significant values
of $\Gamma$ just below $\Omega_{\rm co}$.
This was not the case in previous experiments, {\em e.g.} in \cite{For02}
or in \cite{Mat01}.

In this Letter, we reported the first spectroscopic observation of the
onset of strong scattering of LA waves with $q < q_{\rm co}$
in a network glass.
In that region of $q$ the inhomogeneous linewidth
grows with a very high power of $\Omega$, at least equal to 4, and
probably higher.
Also the intensity shows an anomalous increase with $q$.
It seems that these results could be explained by the resonance
and hybridization of the LA waves with the local BP modes.
If so, these measurements might provide some information about the structure
of glasses at extended length scales about which so little is known otherwise.

Dr. M. Arai is thanked for the excellent sample and Prof. M. Dyakonov for an interesting discussion.


\begin{thebibliography}{}
\bibitem{foot} The spectrum at $Q = 0.97$ nm$^{-1}$ is not usable
owing to a technical problem related to near-forward scattering.
\end{thebibliography}
\end{document}